\documentclass[prl,aps,twocolumn,superscriptaddress]{revtex4}
\usepackage{amsmath}
\usepackage{epsfig}
\usepackage{axodraw}

\newcommand{\insertfig}[2]{\mbox{\epsfxsize=#1cm \epsfbox{#2.eps}}}
\preprint{DOE/ER/40762-322} \preprint{MU-PP\#05-006}

\begin{document}

\title{Fixing Two-Nucleon Weak-Axial Coupling $L_{1,A}$ From $\mu^- d$
Capture}
\author{Jiunn-Wei Chen}
\affiliation{Department of Physics, National Taiwan University, Taipei, Taiwan 10617}
\author{Takashi Inoue}
\affiliation{Department of Physics, National Taiwan University, Taipei, Taiwan 10617}
\affiliation{Department of Physics, Shophia University, Chiyoda-ku, Tokyo 102-8554, Japan}
\author{Xiangdong Ji}
\affiliation{Department of Physics, University of Maryland, College Park, Maryland 20742}
\author{Yingchuan Li}
\affiliation{Department of Physics, University of Maryland, College Park, Maryland 20742}
\date{\today}

\begin{abstract}
We calculate the muon capture rate on the deuteron to
next-to-next-to-leading order in the pionless effective field theory. The
result can be used to constrain the two-nucleon isovector axial coupling $%
L_{1,A}$ to $\pm 2$ fm$^{3}$ if the muon capture rate is measured to 2\%
level. From this, one can determine the neutrino-deuteron break up reactions
and the $pp$-fusion cross section in the sun to a same level of accuracy.
\end{abstract}

\maketitle

The strong evidence of neutrino oscillations observed at the Sudbury
Neutrino Observatory (SNO) \cite{SNO} is based on detecting the $^{8}$B
solar neutrino flux through the following three reactions:
\begin{gather}
\nu _{e}+d\rightarrow p+p+e^{-}~~~(\mathrm{CC}),  \notag \\
\nu _{x}+d\rightarrow p+n+\nu _{x}~~~(\mathrm{NC}),  \notag \\
\nu _{x}+e^{-}\rightarrow \nu _{x}+e^{-}~~~~~~(\mathrm{ES}).
\end{gather}%
The charged current (CC) reaction involves only the electron neutrinos,
while the neutral current (NC) reaction and elastic scattering (ES) involve
all the active neutrinos $\left( x=e,\mu ,\tau \right) $.
The $\nu _{e}$\textbf{\ }and $\nu _{x}$\ fluxes are found to be
significantly different\cite{SNO}.
Further detailed measurements of the fluxes could sharpen the constraints to
neutrino oscillation parameters and provide precision tests to the standard
solar model \cite{SSM}. However, while the ES cross section is known to high
accuracy, the CC and NC cross sections have hadronic uncertainties.
As shown by Butler, Chen and Kong \cite{BCK}, the dominant uncertainties in
low energy CC and NC cross sections comes from the coupling of a two-body
isovector axial current, $L_{1,A}$, in pionless effective field theory ($%
\mathrm{EFT}({\pi \hskip-0.6em/})$). The potential model results of Refs.
\cite{potentialmodel1} and \cite{potentialmodel2} can be reproduced by
different choices of $L_{1,A}$, indicating that the $\sim 5\%$ difference
between the models comes from the different assumptions about the
short-distance nuclear physics. There are other interesting weak reactions
involving the same two-body current, for example, the $pp$ and $pep$ fusion
processes ($pp\rightarrow de^{+}\nu _{e}$, $ppe^{-}\rightarrow d\nu _{e}$)
which power the sun \cite{fusion1,fusion2}. It is one of the great current
interests to measure these neutrino fluxes to further test the standard
solar model.

Recently much effort has been going into determining the effective two-body
axial current interaction \cite{fixing1}. Butler, Chen, and Vogel attempted
to fix $L_{1,A}$ from reactor antineutrino-deuteron breakup reactions, and
they found $L_{1,A}=3.6\pm 5.5~\mathrm{fm}^{3}$ \cite{fixing1}. Chen,
Heeger, and Robertson obtained $L_{1,A}=4.0\pm 6.3~\mathrm{fm}^{3}$ by using
SNO's CC and NC data, calibrated by the ES events of SNO and
Super-Kamiokande(SK) \cite{fixing2}. Schiavilla et al.'s idea \cite{fixing3}
of using the tritium $\beta $ decay rate to control the strength of the
two-body current was adopted by Park et al. in their hybrid EFT calculation
\cite{fix dr}, and the pp fusion rate was predicted with a small error. When
compared with the $\mathrm{EFT}({\pi \hskip-0.6em/})$ calculation \cite%
{fusion2}, their result yields $L_{1,A}=4.2\pm 2.5~$fm$^{3}.$

In this paper we aim to make a high-precision determination of $L_{1,A}$
from the $\mu ^{-}d$ capture process
\begin{equation}
\mu ^{-}+d\rightarrow \nu _{\mu }+n+n \ ,
\end{equation}%
by calculating the rate to next-to-next-to leading order (N$^{2} $LO) in $%
\mathrm{EFT}({\pi \hskip-0.6em/})$. The $\mu ^{-}d$ capture rate has been
measured previously by different groups with rather different results $%
\Gamma ^{\mathrm{exp}}=470\pm 29~s^{-1}$ \cite{mu-d-exp1} and $\Gamma ^{%
\mathrm{exp}}=409\pm 40~s^{-1}$ \cite{mu-d-exp2}. A measurement of this rate
with 1\% precision is under investigation at PSI \cite{Kammel}. An earlier
potential model calculation \cite{tkk} gave $\Gamma =397\sim 400~s^{-1}$.
More recently, the hybrid approach mentioned above gave $\Gamma =386\pm
5~s^{-1}$\ \cite{Ando}.

A concern in applying $\mathrm{EFT}({\pi \hskip-0.6em/})$ to the $\mu ^{-}d $
capture is that the energy transfer into the hadronic system might be too
large to apply $\mathrm{EFT}({\pi \hskip-0.6em/})$. However, as shown in Ref.%
\cite{Ando} and also in this calculation, the contribution to the total rate
from high-energy neutrons is small, and it is possible to impose a neutron
energy cut to isolate the low-energy ($\leq 10-20$ MeV) neutron events
without significantly increasing the statistical errors \cite{fixing1,Kammel}%
.


Effective field theory is useful when low and high energy scales in the
problem are widely separated. For low-energy processes, short-distance
physics can be taken into account by local operators in an effective
lagrangian involving only low-energy degrees of freedom. For $\nu (\bar{\nu}%
)-d$ scattering with neutrino energy below 20 MeV and $\mu ^{-}d$ capture
with small final-state neutron energy, the pion and other meson exchanges
are not dynamical degrees of freedom, and their physics can be captured by
contact interactions involving nucleons and the external currents. To make
predictions with controlled precision, calculations are done with the
perturbative expansion parameter $Q\equiv (1/a_{nn}^{(^{1}S_{0})},\gamma
,p)/\Lambda \ ,$ which is the ratio of light to heavy scales. The light
scales include the inverse S-wave neutron-neutron scattering length $%
1/a_{nn}^{(^{1}S_{0})}=-10.6$\textrm{\ }MeV in the $^{1}S_{0}$ channel, the
deuteron binding momentum $\gamma =45.7$ MeV in the $^{3}S_{1}$ channel, and
typical nucleon momentum $p$ in the system. The heavy scale $\Lambda $ is
set by the pion mass $m_{\pi }$. This $\mathrm{EFT}({\pi \hskip-0.6em/})$
(see e.g. \cite{pionless}) and its dibaryon version \cite%
{dibaryon1,dibaryon2,dibaryon3} have been applied successfully to
many processes involving the deuteron, including electro-magnetic
processes such as Compton
scattering $\gamma d\rightarrow \gamma d$ \cite{Compton1,Compton2}, $%
np\rightarrow d\gamma $ relevant to the big-bang nucleosynthesis \cite%
{Nsynthesis1,Nsynthesis2}, weak processes such as $\nu d$ reactions for SNO
physics\cite{BCK}, the solar $pp$ fusion process \cite{fusion1,fusion2}, and
parity violating observables \cite{pv}. For reviews on three-body systems,
see \cite{3body}.

The effective Lagrangian for the CC weak interaction is given by $\mathcal{L}%
^{CC}=-\sqrt{\omega _{i}}G_{F}V_{ud}l_{+}^{\mu }J_{\mu }^{-}/\sqrt{2}+%
\mathrm{h.c.},$ where $G_{F}=1.166\times 10^{-5}\mathrm{GeV}^{-2}$ is the
Fermi coupling constant and $\omega _{i}=1.024$\textbf{\ }takes into account
the inner electroweak radiative correction \cite{radiative}. $l_{+}^{\mu }=%
\overline{\nu }_{\mu }\gamma ^{\mu }(1-\gamma _{5})\mu $ is the leptonic
current. The quark current $J_{\mu }^{-}=V_{\mu }^{-}-A_{\mu }^{-}=(V_{\mu
}^{1}-A_{\mu }^{1})-i(V_{\mu }^{2}-A_{\mu }^{2})$ contains both vector and
axial-vector interactions, where the superscripts 1 and 2 are the isospin
indices. At the scale relevant to nuclear physics, the quark current need be
matched to a hadronic current which in general contains one-nucleon,
two-nucleon, etc., operators.

Up to the order of our interest, the one-nucleon isovector vector and axial
vector currents are
\begin{eqnarray}
V_{(1)}^{0,a} &=&N^{\dagger }\frac{\tau ^{a}}{2}\left( 1+\frac{1}{6}\langle
r^{2}\rangle _{c}^{I=1}\overline{\nabla }^{2}\right) N\,  \notag \\
V_{(1)}^{k,a} &=&\frac{i}{2M_{N}}N^{\dagger }\overleftrightarrow{\nabla }_{k}%
\frac{\tau ^{a}}{2}N-\frac{\mu ^{(1)}}{M_{N}}\epsilon _{kij}N^{\dagger
}\sigma _{i}\overline{\nabla }_{j}\frac{\tau ^{a}}{2}N\ ,  \notag \\
A_{(1)}^{0,a} &=&\frac{ig_{A}}{2M_{N}}N^{\dagger }{\vec{\sigma}}\cdot
\overleftrightarrow{\nabla }\frac{\tau ^{a}}{2}N \\
&&-\frac{1}{4M_{N}^{2}}N^{\dagger }G_{p}\left( \overline{\nabla }^{2}\right)
\overline{\partial _{0}}{\vec{\sigma}}\cdot \overline{\nabla }\frac{\tau ^{a}%
}{2}N\ ,  \notag \\
A_{(1)}^{k,a} &=&{g_{A}}N^{\dagger }\mathbf{\sigma }_{k}\frac{\tau ^{a}}{2}%
\left( 1+\frac{1}{6}\langle r^{2}\rangle _{A}^{I=1}\overline{\nabla }%
^{2}\right) N  \notag \\
&&+\frac{1}{4M_{N}^{2}}N^{\dagger }G_{p}\left( \overline{\nabla }^{2}\right)
\overline{\nabla }_{k}{\vec{\sigma}}\cdot \overline{\nabla }\frac{\tau ^{a}}{%
2}N\ ,  \notag
\end{eqnarray}%
where\textbf{\ }$\overline{\nabla }=\overleftarrow{\nabla }+\overrightarrow{%
\nabla }$\textbf{, }$\overline{\partial _{0}}=\overleftarrow{\partial _{0}}+%
\overrightarrow{\partial _{0}}$\textbf{, }and\textbf{\ }$\overleftrightarrow{%
\nabla }=\overleftarrow{\nabla }-\overrightarrow{\nabla }.$\textbf{\ }The
superscript $a$ is the isospin index, $\mu ^{(1)}=(\mu _{p}-\mu _{n})/2=2.353
$, is the isovector magnetic moment. Isovector Dirac charge radius $\langle
r^{2}\rangle _{c}^{I=1}=\langle r^{2}\rangle _{c}^{p}-\langle r^{2}\rangle
_{c}^{n}=0.873~$fm$^{2}$, and the isovector axial-charge radius $\langle
r^{2}\rangle _{A}^{I=1}\simeq 0.45~\mathrm{fm}^{2}$. We have neglected terms
of order $p^{2}/M_{N}^{2}$ or even $\mu ^{(1)}p^{2}/M_{N}^{2}$. The
pseudoscalar form factor is, to a good approximation, dominated by the
pion-pole $G_{p}(q^{2})=\frac{4M_{N}^{2}g_{A}}{M_{\pi }^{2}-q^{2}}$ whose $%
q^{2}$ dependence will not be expanded because the momentum transfer $|%
\mathbf{q}|$ is of order muon mass $M_{\mu }$\ with low energy final state
neutrons. The $G_{p}$ contribution to the axial current is counted of order $%
Q$. 

The lowest dimensional two-nucleon isovector currents, in the dibaryon
version of $\mathrm{EFT}({\pi \hskip-0.6em/})$, relevant to the $\mu ^{-}d$
capture process are
\begin{eqnarray}
V_{(2)}^{k,a} &=&\frac{L_{1}^{db}}{M_{N}\sqrt{%
r^{(^{3}S_{1})}r_{nn}^{(^{1}S_{0})}}}\epsilon _{kij}t_{i}^{\dagger }%
\overline{\nabla }_{j}s_{a}+\mathrm{h.c.}\ ,  \notag \\
A_{(2)}^{k,a} &=&\frac{L_{1,A}^{db}}{M_{N}\sqrt{%
r^{(^{3}S_{1})}r_{nn}^{(^{1}S_{0})}}}\left( t_{k}^{\dagger }s_{a}\right.  \\
&&\left. +\frac{G_{p}}{4M_{N}^{2}g_{A}}t_{j}^{\dagger }\overline{\nabla }_{j}%
\overline{\nabla }_{k}s_{a}\right) +\mathrm{h.c.},  \notag
\end{eqnarray}%
where $t_{i}$ and $s_{a}$ are dibaryon fields for the two-nucleon $^{3}S_{1}$
and $^{1}S_{0}$ states, respectively. The second term in $A_{(2)}^{k,a}$ is
induced by the $G_{p}$ term in the one-nucleon current. $%
r^{(^{3}S_{1})}=1.764~\mathrm{fm}$ and $r_{nn}^{(^{1}S_{0})}=2.8~\mathrm{fm}$
are the effective ranges in triplet and two-neutron singlet channels,
respectively. $L_{1}^{db}$ and $L_{1,A}^{db}$ are coupling constants in
dibaryon formalism. The vector current is N$^{2}$LO and its coupling $%
L_{1}^{db}=-4.08~\mathrm{fm}$ has been determined by the rate of $%
n+p\rightarrow d+\gamma $ near threshold. The axial current is NLO, and its
coupling $L_{1,A}^{db}$ is proportional to the renormalization-scale-$\mu $%
-independent $\tilde{L}_{1,A}$ in Ref. \cite{BCK} as $L_{1,A}^{db}=\frac{%
M_{N}}{2\pi }\tilde{L}_{1,A}$, through which $L_{1,A}^{db}$ is related to
the $\mu $-dependent $L_{1,A}(\mu )$ in Ref. \cite{BCK}. The numerical
relation between $L_{1,A}^{db}$ and $L_{1,A}(\mu )$ is $%
L_{1,A}^{db}=-13.8+0.28L_{1,A}(M_{\pi })$, where $L_{1,A}(M_{\pi })$ is in
units of $\mathrm{fm}^{3}$ and has a natural size $\sim 6$ fm$^{3}$.

The $\mu ^{-}d$ atom has a ground state with a hyperfine structure,
corresponding to the total angular momentum $F=3/2$ and $F=1/2$. The $\mu
^{-}d$ capture process is known to take place almost uniquely from the
doublet $F=1/2$ state. The differential capture rate for muon and deuteron
in their specific polarization states can be written in terms of leptonic
tensor $l^{\mu \nu }$ and hadronic tensor $W_{\mu \nu }$ as
\begin{equation}
\frac{d\Gamma (S,\hat{\xi})}{dEd\Omega }=\frac{\omega
_{i}G_{F}^{2}|V_{ud}|^{2}E|\psi (0)|^{2}}{32\pi ^{2}M_{\mu }(1+\frac{M_{\mu }%
}{M_{d}})}l^{\mu \nu }(S)W_{\mu \nu }(\hat{\xi}),
\end{equation}%
where the $|\psi (0)|^{2}=\frac{1}{\pi }\left( \alpha _{\mathrm{em}}\frac{%
M_{\mu }M_{d}}{M_{\mu }+M_{d}}\right) ^{3}$ is the $1S$-state
wave-function-at-origin-squared, $E$ is the outgoing neutrino energy, and $%
M_{d}$ is the mass of the deuteron.

The capture rate depends on the polarization vector of the muon $S_{\mu }$
and the deuteron polarization vector $\hat{\xi}$. The leptonic tensor is
given by $l^{\mu \nu }(S)=4(k^{\mu }k^{\prime }{}^{\nu }+k^{\prime }{}^{\mu
}k^{\nu }-k\cdot k^{\prime }g^{\mu \nu }+i\epsilon ^{\mu \nu \rho \sigma
}k_{\rho }k_{\sigma }^{\prime })-4M_{\mu }(S^{\mu }k^{\prime }{}^{\nu
}+k^{\prime }{}^{\mu }S^{\nu }-k^{\prime }\cdot Sg^{\mu \nu }+i\epsilon
^{\mu \nu \rho \sigma }S_{\rho }k_{\sigma }^{\prime })$, where $k=(M_{\mu },%
\vec{0})$ and $k^{\prime }=(E,\vec{k^{\prime }})$, with $E=|\vec{k^{\prime }}%
|$, are the four-momenta of initial muon and final $\nu _{\mu }$,
respectively. The hadronic tensor is
\begin{equation}
W_{\mu \nu }(\hat{\xi})=\frac{1}{\pi }\mathrm{Im}\left[ \int
d^{4}xe^{iqx}T\langle d|J_{\mu }^{+}(x)J_{\nu }^{-}(0)|d\rangle \right] ,
\end{equation}%
where $q_{\mu }=k_{\mu }-k_{\mu }^{\prime }$, and $|d\rangle \equiv |d(P,%
\hat{\xi})\rangle $ is the deuteron state with momentum $P=(M_{d},\vec{0})$
and polarization $\hat{\xi}$.



The diagrams contributing to $W_{\mu \nu }(\hat{\xi})$ up to N$^{2}$LO are
shown in Fig. 1. A straightforward calculation finds
\begin{eqnarray}
&&\frac{d\Gamma }{dE}=\frac{G_{F}^{2}|V_{ud}|^{2}E^{2}|\psi (0)|^{2}}{2\pi
(1+\frac{M_{\mu }}{M_{d}})}\frac{1}{1-\varepsilon \gamma r^{(^{3}S_{1})}}%
\left[ \left( F_{1}-\frac{F_{2}}{2}\right) \right.   \notag \\
&&\left( 3G_{V}^{2}-2(G_{V}-G_{A})^{2}-\varepsilon \frac{4g_{A}(1-g_{A})M_{%
\mu }E}{3(M_{\pi }^{2}-q^{2})}\right.   \notag \\
&&\left. +\varepsilon \kappa ^{(1)}\frac{8(1-g_{A})E}{3M_{N}}\right) +\left(
F_{1}-\frac{F_{2}}{6}+\frac{2}{3}F_{4a}\right) \left( 9G_{A}^{2}\right.
\notag \\
&&\left. -\varepsilon \frac{6g_{A}^{2}M_{\mu }E}{M_{\pi }^{2}-q^{2}}%
+\varepsilon ^{2}\frac{g_{A}^{2}E^{4}}{\left( M_{\pi }^{2}-q^{2}\right) ^{2}}%
\right)   \notag \\
&&+\left( 10F_{1}-F_{2}+8F_{4b}\right) \frac{\varepsilon ^{2}\kappa
^{(1)^{2}}E^{2}}{3M_{N}^{2}}  \notag \\
&&+\left( F_{1}-\frac{F_{2}}{6}+\frac{2}{3}F_{4c}\right) 12\varepsilon
g_{A}\kappa ^{(1)}\frac{E}{M_{N}}  \notag \\
&&-\left( F_{1}+F_{4c}\right) \frac{8\varepsilon ^{2}g_{A}\kappa ^{(1)}E^{3}%
}{3M_{N}(M_{\pi }^{2}-q^{2})}  \notag \\
&&\left. +F_{5}\varepsilon ^{2}E\left( \frac{14}{3}g_{A}^{2}+\frac{16}{3}%
g_{A}+2\right) \right]
\end{eqnarray}%
where $G_{V}=1+\varepsilon ^{2}\frac{1}{6}\langle r^{2}\rangle
_{c}^{I=1}q^{2}$ is the Dirac form factor, and $G_{A}=g_{A}(1+\varepsilon
^{2}\frac{1}{6}\langle r^{2}\rangle _{c}^{I=1}q^{2})$ is the axial form
factor. The $\varepsilon $ is formally introduced to keep track of the $Q$
expansion. After expanded in $\varepsilon $ and truncated at $\varepsilon
^{2}$, the N$^{2}$LO result is obtained by setting $\varepsilon =1$. The
functions $F_{1a,1b}$, $F_{2}$, and $F_{3a,3b,3c}$ are from diagrams $(a)$, $%
(b)$, and $(c)$, respectively, in Fig. 1
\begin{eqnarray}
F_{1a} &=&\frac{2M_{N}p\gamma }{\pi (M_{N}^{2}\Delta _{E}^{2}-p^{2}E^{2})}
\notag \\
F_{1b} &=&\frac{\gamma }{\pi E^{3}}\left[ \frac{2M_{N}Ep\Delta _{E}}{%
M_{N}^{2}\Delta _{E}^{2}-p^{2}E^{2}}\right. \left. +\ln \left( \frac{%
M_{N}\Delta _{E}-pE}{M_{N}\Delta _{E}+pE}\right) \right]   \notag \\
F_{2} &=&\frac{4\gamma }{\pi \Delta _{E}E}\mathrm{tanh}^{-1}\left( \frac{Ep}{%
M_{N}\Delta _{E}}\right)   \notag \\
F_{3a} &=&\frac{2\gamma M_{N}^{2}}{\pi ^{2}E^{2}}\mathrm{Im}\left\{
G_{1}^{2}A^{(^{1}S_{0})}(p)\right\}   \notag \\
F_{3b} &=&\frac{2\gamma M_{N}^{2}}{\pi ^{2}E^{2}}\mathrm{Im}\left\{
G_{2}^{2}A^{(^{1}S_{0})}(p)\right\}   \notag \\
F_{3c} &=&\frac{2\gamma M_{N}^{2}}{\pi ^{2}E^{2}}\mathrm{Im}\left\{
G_{1}G_{2}A^{(^{1}S_{0})}(p)\right\} ,
\end{eqnarray}%
with the re-scattering amplitude in the singlet channel as $%
A^{(^{1}S_{0})}(p)=-\frac{4\pi }{M_{N}}(\frac{1}{a_{nn}^{(^{1}S_{0})}}%
-\varepsilon \frac{1}{2}r_{nn}^{(^{1}S_{0})}p^{2}+ip)^{-1}$, and functions $%
G_{1,2}$ are defined as $G_{1}=\mathrm{tan}^{-1}(\frac{E}{2(\gamma -ip)}%
)+\varepsilon \frac{E}{4M_{N}g_{A}}L_{1,A}^{db}$ and $G_{2}=\mathrm{tan}%
^{-1}(\frac{E}{2(\gamma -ip)})+\varepsilon \frac{E}{4\mu ^{(1)}}L_{1}^{db}$.
The energy injection into the two-neutron system is $\Delta _{E}\equiv
M_{\mu }-E-(M_{n}-M_{p})$. The relative momentum between the two final-state
neutrons is $2p$, with $p=\sqrt{M_{N}\Delta _{E}-\gamma ^{2}-E^{2}/4}$.
\begin{figure}[t]
\begin{center}
\mbox{
\begin{picture}(0,50)(100,0)

\put(-15,-10){\insertfig{8.0}{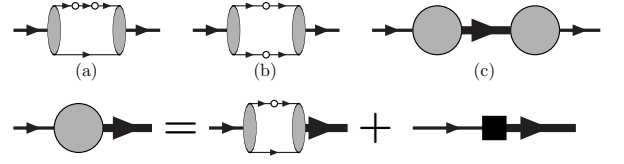}}

\end{picture}
}
\end{center}
\caption{Diagrams contributing to $W_{\protect\mu \protect\nu }(\hat{\protect%
\xi})$ up to N$^{2}$LO. The lines inside the loop are the nucleon
propagators, and the thin and thick lines outside represent the triplet and
singlet dibaryon fields, respectively. The small open circle represents the
insertion of hadronic one-body current. In diagram (c), the large gray
circle indicates the two possible hadronic current insertions shown in the
second row where the dark square denotes the two-body current associated
with $L_{1,A}$ and $L_{1}$.}
\end{figure}
\begin{figure}[t]
\begin{center}
\mbox{
\begin{picture}(0,130)(100,0)

\put(8,-10){\insertfig{7.8}{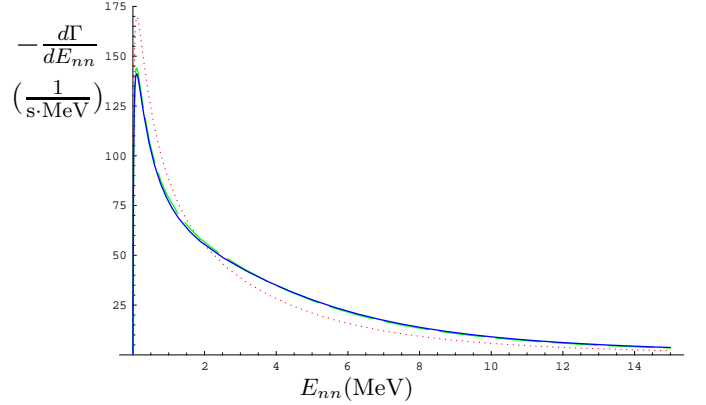}}

\Text(105,-15)[c]{{$E_{nn}$(MeV)}}

\Text(-8,115)[c]{\large{$-\frac{d\Gamma}{dE_{nn}}$}}

\Text(-8,95)[c]{\large{$(\frac{1}{{\rm s \cdot MeV}})$}}

\end{picture}
}
\end{center}
\caption{(Color online) The differential capture rate $d\Gamma /dE_{nn}$ calculated using $%
L_{1,A}=6~\mathrm{fm}^{3}$. The doted line is the LO result. The
dashed and
solid lines, which sit on top of each other in this scale, are the NLO and N$%
^{2}$LO results, respectively.}
\end{figure}




Power counting for the present calculation is rather tricky, and
it would be misleading, for example, to just use the $p/m_{\pi }$
to estimate the accuracy of the expansion. For example, it is
well-know that in the nucleon-nucleon scattering, the effective
theory without pion works rather well at the nucleon momentum on
the order of 100 MeV, a value close to the pion mass. This can
also be seen in the present calculation because the NLO result at
large neutron momentum does not completely modify the leading
order result, whereas the naive power counting would indicate
otherwise. Fortunately, for muon capture, the most of the events
occur when the neutron momentum is about half of the muon mass,
that is, about 50 MeV.

In Fig. 2 we show the differential rate $d\Gamma /dE_{nn}$ in terms of the
relative motion energy $E_{nn}=2(\sqrt{M^2_N+p^2}-M_N)$ of two final-state
neutrons in the region where the $\mathrm{EFT}({\pi \hskip-0.6em/})$
calculation is most reliable. It is clear from the figure that the
differential rate in the energy region $E_{nn}\geq 10$\ MeV is very small,
and is negligible for $E_{nn}\geq 15$\ MeV. By comparing the results of LO,
NLO, and N$^{2}$LO, we find good convergence of the expansion.

In the case that a neutron energy cut can be imposed on experimental data
\cite{Kammel}, it is possible to define and measure the integrated capture
rate up to a threshold energy $E_{nn}^{\mathrm{th}}$
\begin{equation}
\Gamma (E_{nn}^{\mathrm{th}})\equiv -\int_{0}^{E_{nn}^{\mathrm{th}}}\frac{%
d\Gamma }{dE_{nn}}dE_{nn}.
\end{equation}%
The result up to 30 MeV is shown in Fig. 3. In the whole energy region, the
NLO contribution is less than $10\%$ of the LO contribution, while the N$^{2}
$LO contribution is less than $1\%$. This small size of N$^{2}$LO
contribution is accidental and does not happen for unpolarized and $F=3/2$
rates. For example, the unpolarized rate has the expansion (for $%
L_{1,A}=6$ fm$^{3}$)
\begin{equation}
\Gamma _{unpol.}=\Gamma _{unpol.}^{LO}(1+5.3\%+4.9\%),
\end{equation}
where the NLO correction is abnormally small, and NNLO is of the
normal size.
A similar expansion is obtained for $L_{1,A}=-6$ fm$^{3}$:%
\begin{equation}
\ \Gamma _{unpol.}=\Gamma _{unpol.}^{LO}(1+10.9\%+5.2\%)
\end{equation}
which shows a nice convergence pattern. Based on this trend, we
assign a 2-3\% correction at NNNLO, corresponding to an error in
$L_{1,A}\sim 2~\mathrm{fm}^{3}$. This is consistent with the naive
estimation of 3\% if the small expansion parameter is $1/3$. A
calculation shows the N$^{3}$LO final state P-wave re-scattering
contributes only $\sim 1\%$. Furthermore, the result is
insensitive to the uncertainty in $a_{nn}^{(^{1}S_{0})}$. Choosing
$L_{1,A}=5.6~\mathrm{fm}^{3}$, the energy dependence of our result
matches the previous hybrid calculation very well \cite{Ando}.
\begin{figure}[t]
\begin{center}
\mbox{
\begin{picture}(0,130)(100,0)

\put(2,-8){\insertfig{8}{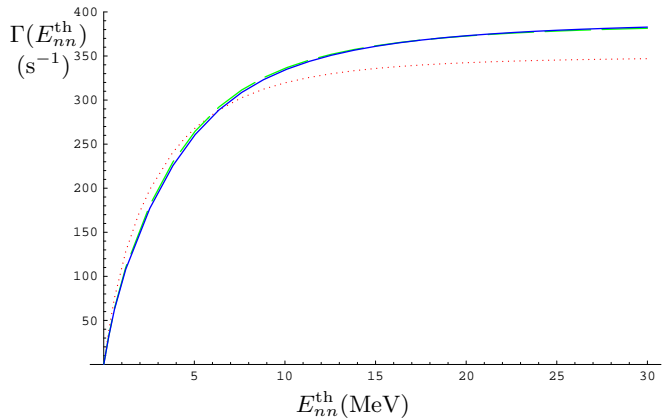}}

\Text(110,-15)[c]{{$E_{nn}^{\rm th}$(MeV)}}

\Text(-5,125)[c]{{$\Gamma(E_{nn}^{\rm th})$}}
\Text(-7,113)[c]{$({\rm s}^{-1})$}
\end{picture}
}
\end{center}
\caption{(Color online) The integrated capture rate $\Gamma
(E_{nn}^{\mathrm{th}})$
calculated using $L_{1,A}=6~\mathrm{fm}^{3}$ for the relative energy $%
E_{nn}^{\mathrm{th}}$ of the two neutrons up to 30 MeV. The description of
the lines is the same as that in the caption of Fig. 2.}
\end{figure}

To extract $L_{1,A}$ from experimental data, it is useful to provide the
dependence of the rate on $L_{1,A}$
\begin{equation}
\Gamma (E_{nn}^{\mathrm{th}})=a(E_{nn}^{\mathrm{th}})+b(E_{nn}^{\mathrm{th}%
})L_{1,A},
\end{equation}%
where $L_{1,A}$ is in unit of $\mathrm{fm}^{3}$. The energy-cut dependent
functions $a(E_{nn}^{\mathrm{th}})$ and $b(E_{nn}^{\mathrm{th}})$ for a set
of $E_{nn}^{\mathrm{th}}$s are listed in the Table 1, from which we observe
that, for the whole range of $E_{nn}^{\mathrm{th}}$, the size of $b(E_{nn}^{%
\mathrm{th}})$ is about $1.3\sim 1.5\%$ of the size of $a(E_{nn}^{\mathrm{th}%
})$. This shows how an error in capture rate is translated into an
uncertainty of $L_{1,A}$. \bigskip \newline
\centerline{Table 1:} \newline
Coefficients functions $a(E_{nn}^{\mathrm{th}})$ and $b(E_{nn}^{\mathrm{th}%
}) $ for specific values of two-neutron relative energy $E_{nn}^{\mathrm{th}%
} $ from the $\mathrm{EFT}({\pi \hskip-0.6em/})$ calculation. \bigskip
\newline
\begin{tabular}{rrrrr}
\hline
$E_{nn}^{\mathrm{th}}(\mathrm{MeV})$ & ~~~~5.0 & ~~~~10.0 & ~~~~15.0 &
~~~~20.0 \\ \hline
$a(E_{nn}^{\mathrm{th}})(\mathrm{s}^{-1})$ & ~~~~239.2 & ~~~~308.0 &
~~~~332.0 & ~~~~342.3 \\ \hline
$b(E_{nn}^{\mathrm{th}})(\mathrm{s}^{-1}\mathrm{fm}^{-3})$ & ~~~~3.3 &
~~~~4.2 & ~~~~4.7 & ~~~~4.9 \\ \hline
\end{tabular}

\bigskip

In summary, we calculated the $\mu ^{-}d$ capture rate using $\mathrm{EFT}({%
\pi \hskip-0.6em/})$ to N$^{2}$LO. The major goal is to fix the two-nucleon
isovector axial coupling constant $L_{1,A}$ from future precision
experimental data. An experimental result on the integrated rate up to some
neutron energy $E_{nn}^{\mathrm{th}}$ with a $2\%$ error should be able to,
through comparison with our calculation with theoretical error 2-3$\%$, fix
the $L_{1,A}$ with error $\pm 2.0~\mathrm{fm}^{3}$. This in turn allows us
to determine the neutrino deuteron breakup cross section and the pp fusion
rate in the sun to 2-3\%.

This work was supported by the U. S. Department of Energy via grant
DE-FG02-93ER-40762 and the NSC of Taiwan. We thank S. Ando, D.W. Hertzog, P.
Kammel, K. Kubodera, and T.-S. Park for helpful discussions.

\appendix

\end{document}